\documentclass[twoside]{IEEEtran}

\usepackage[english]{babel}
\usepackage[dvips]{graphicx}
\usepackage{subfigure}
\usepackage{amsmath}
\usepackage{amssymb}

\newcommand{\tcq}{trellis-coded quantization}

\newcommand{\tcqr}{trellis-coded quantizer}

\newcommand{\dsc}{distributed source coding}
\newcommand{\DSC}{Distributed Source Coding}
\newcommand{\cvs}{continuous-valued syndrome}
\newcommand{\CVS}{Continuous-Valued Syndrome}
\newcommand{\csi}{coding with side information}
\newcommand{\CSI}{Coding with Side Information}

\begin{document}
\title{Distributed Source Coding Using Continuous-Valued Syndromes}

\author{Lorenzo~Cappellari,~\IEEEmembership{Member,~IEEE}%
\thanks{L.~Cappellari is with the Dept.~of Information Engineering of the University of Padova, Italy.}}

\maketitle

\begin{abstract}
This paper addresses the problem of coding a continuous random source correlated with another source which is only available at the decoder. The proposed approach is based on the extension of the channel coding concept of syndrome from the discrete into the continuous domain. If the correlation between the sources can be described by an additive Gaussian backward channel and capacity-achieving linear codes are employed, it is shown that the performance of the system is asymptotically close to the Wyner-Ziv bound. Even if such an additive channel is not Gaussian, the design procedure can fit the desired correlation and transmission rate. Experiments based on trellis-coded quantization show that the proposed system achieves a performance within 3-4 dB of the theoretical bound in the 0.5-3 bit/sample rate range for any Gaussian correlation, with a reasonable computational complexity.
\end{abstract}


\IEEEpeerreviewmaketitle

\section{Introduction}
\PARstart{D}{istributed source coding} addresses the problem of coding the outcomes of multiple correlated sources independently, i.e.~without allowing the respective encoders to communicate with each other. Theoretical results on this topic appeared in the seventies, and showed that, if the joint distribution characterizing the correlation structure is known, there is no loss in performance with respect to (w.r.t.) the case where the encoders can collaborate with each other. This result goes under the name of the Slepian-Wolf coding theorem \cite{cover_ElemInfoThNew,slepian_noiselessCodingCorrSrc}. A related problem is {\csi} at the decoder, in which the outcomes of a source have to be encoded and decoded within a given distortion under the condition that a second correlated source is only available at the decoder. Theoretical performance bounds, given by Wyner and Ziv \cite{wyner_RDFunctSrcCodingSideInfo}, show that in certain cases there is no loss in performance w.r.t.~the case where the encoder can access the side information as well.

Coding methods that approach these theoretical limits are hence highly desirable, and have many practical applications. For example, they can be employed in sensor networks \cite{xiong_DSCSensors}, where communication between the nodes not only requires an elaborate intersensor network, but also may be limited by bandwidth constraints. In addition to this natural {\dsc} application, these coding methods can be also used in video coding, not only for independent encoding of multiple correlated video sensors \cite{flierl_multiview}, but mostly to reduce the encoding complexity w.r.t.~the classical, non-distributed, video coding solutions \cite{girod_DVC,puri_DVC}. Hyperspectral image compression is yet another field where {\dsc} principles can be taken into account \cite{tang_hyperspecWD-DSC,nonnis_hyperSpectralSWCoding,cheung_corrEstHyperspecDSC}. 

Despite the limits of {\dsc} were theoretically investigated more than thirty years ago, practical solutions to the problem have been proposed only in the last decade. All solutions stemmed from the connection that {\dsc} has to channel coding, as already pointed out by Wyner in its pioneering work \cite{wyner_resultsShannon}. Algorithms based on turbo and low-density parity-check (LDPC) codes \cite{berrou_turboJCOM,mackay_ldpc} have then appeared that approach the theoretical bounds in the discrete binary domain \cite{aaron_compressionSideTurboCodes,liveris_sideLDPC}.

To solve the {\csi} issue for continuous random sources, current proposals recast the problem into a binary domain or, at least, into a discrete domain, by using some kind of quantization. For example, in the \emph{{\dsc} using syndromes} (DISCUS) system \cite{pradhan_DISCUS1,pradhan_DISCUS2,pradhan_genCosetCodesBinning}, the random variables are transformed into a discrete domain, in which cosets of some trellis codes are identified on top of the reconstruction codebook used by the quantizer, before actually coding them. The related idea of using nested codes for {\dsc} is discussed by Zamir \emph{et al.}~in \cite{zamir_nestedCodesWZencoding,zamir_nestedCodesStructBin}.

In this work it is proposed to solve the {\csi} problem for continuous random sources entirely in the continuous domain. By operating in this domain, the problem is recast into the traditional source coding problem of a \emph{\cvs} that is closely related to the actual statistical correlation between the source and the side information. In a certain sense, the paper is an extension of the work in \cite{zamir_nestedCodesWZencoding}, since it is shown that, under some condition, it is not necessary for the codes to be nested for optimum performance, as first proved in \cite{cappellari_DSCcontinueSyndrome}.

The rest of the paper is organized as follows. In Section \ref{s:math_review}, a review on linear block codes and lattices is given, and the concept of \emph{syndrome} is introduced. Section \ref{s:dsc_review} discusses the duality between the problems of {\dsc} and of channel coding. The discussion is presented from a perspective that enables us to directly utilize the concepts introduced in Section \ref{s:math_review} to solve the {\dsc} problem. This approach is presented in Section \ref{s:algorithm}, and the experimental results are discussed in Section \ref{s:experiments}. Conclusions are drawn in Section \ref{s:conclusion}.

\section{Linear Block Codes and Lattices}\label{s:math_review}
In this section, the basic properties of linear block codes and lattices are reviewed. In particular, the focus is on the concept of \emph{syndrome}. After reading this section, it should be clear that the definition of such a concept catches some dualities between the two algebraic structures.

\subsection{Linear Block Codes and Syndromes}\label{s:syndrome_review}
Consider the Galois field $\mathbb{F}=GF(q)$, with $q=p^m$ for some prime number $p\geq 2$ and some $m\in\{1,2,\dots\}$. An $(n,k)$-\emph{linear block code} $\mathcal{C}$ over $\mathbb{F}$ is a $k$-dimensional subspace of the vector space $\mathbb{F}^n$ over $\mathbb{F}$. Equivalently,
\begin{equation}
\label{e:block_code}
\mathcal{C} \triangleq \left\{ \sum_{i=1}^k a_ig_i: a_i\in\mathbb{F}, \;\forall i=1,2,\dots,k\right\}\;,
\end{equation}
for some set $G=\{g_i\}_{i=1,2,\dots,k}$ of \emph{linearly independent} vectors on $\mathbb{F}^n$. It is worth noting that an $(n',k',m)$-\emph{convolutional code} over $\mathbb{F}$, for a sufficiently large number $L$ of consecutive $n'$-length output codewords, can be as well described by (\ref{e:block_code}), with $n=n'L$ and $k=k'L-m$ \cite{mceliece_ConvCodes_inbook}.

Being a subgroup of the additive group $(\mathbb{F}^n,+)$, $\mathcal{C}$ induces a partition of $\mathbb{F}^n$ into cosets. In particular, the set of all cosets is the quotient group $\mathbb{F}^n/\mathcal{C}$, and, since the cardinality of $\mathcal{C}$ is $|\mathcal{C}|=q^k$, there are exactly $|\mathbb{F}^n|/|\mathcal{C}|=q^n/q^k=q^{n-k}$ of them. To identify the coset to which an element of $\mathbb{F}^n$ belongs, consider any linear application of $\mathbb{F}^n$ with kernel equal to $\mathcal{C}$ into some other $(n-k)$-dimensional vector space over $\mathbb{F}$. Then, such a homomorphism of $(\mathbb{F}^n,+)$ automatically assigns a distinctive vector, called \emph{syndrome}, to the elements belonging to the same coset.

A simple way to construct such a linear application is as follows. Consider the orthogonal complement $\mathcal{C}^\perp$ of $\mathcal{C}$ into $\mathbb{F}^n$ (i.e.~the \emph{dual code} of $\mathcal{C}$, with the canonical inner product on $\mathbb{F}^n$). Since $\mathcal{C}^\perp$ is an $(n-k)$-dimensional subspace of $\mathbb{F}^n$, it is generated by some set $H=\{h_i\}_{i=1,2,\dots,n-k}$ of linearly independent vectors on $\mathbb{F}^n$. Associate with each $a\in\mathbb{F}^n$ the vector $s_H(a)$ on $\mathbb{F}^{n-k}$ whose canonical coordinates are the inner products $\langle a,h_i\rangle$, $i=1,2,\dots,n-k$. This vector represents the syndrome of $a$ relative to the $(n-k)\times n$ \emph{parity check matrix} $\mathbf{H}$ (whose rows contain the canonical coordinates of each $h_i$). In addition, it is in principle straightforward to compute $s_H(a)$ since it can be simply obtained by matrix multiplication with $\mathbf{H}^T$.

The syndrome plays an important role in minimum-distance decoding (using some distance function such as the Hamming distance) when it can be univocally associated with a minimum-weight element in each coset of $\mathbb{F}^n/\mathcal{C}$. In this case the minimum-weight element that by subtraction leads to a codeword of $\mathcal{C}$ can be identified from the syndrome of any received \emph{codeword}. Linear codes with good distance properties permit to associate a unique minimum-weight element with most of the syndromes. When a syndrome without this property occurs, an \emph{error} is \emph{detected} but not \emph{corrected}.

\subsection{Lattices and {\CVS}s}\label{s:c_syndrome_review}
Consider now the additive group of real numbers $(\mathbb{R},+)$. An $n$-dimensional \emph{lattice} $\Lambda$ is a \emph{discrete} subgroup of the Euclidean space $\mathbb{R}^n$ that spans $\mathbb{R}^n$ itself. The group $(\Lambda,+)$ can be equivalently defined as
\begin{equation}
\Lambda \triangleq \left\{ \sum_{i=1}^n a_iv_i: a_i\in\mathbb{Z}, \;\forall i=1,2,\dots,n\right\}\;, \nonumber
\end{equation}
for some set $\{v_i\}_{i=1,2,\dots,n}$ of \emph{linearly independent} vectors on $\mathbb{R}^n$.

Through the quotient group $\mathbb{R}^n/\Lambda$, any lattice induces a partition of $\mathbb{R}^n$ into cosets. For practical reasons, it is useful to identify each coset using one of its elements. Hence, consider an injection $l:\mathbb{R}^n/\Lambda \to \mathbb{R}^n$ such that $l(A)\in A$, $\forall A\in\mathbb{R}^n/\Lambda$, and call \emph{labeling function} of $\mathbb{R}^n/\Lambda$ any such function. The \emph{fundamental region} $R_l(\Lambda)$ of $\Lambda$ \emph{induced by} $l$ is then defined as the image of $l$, i.e.~$R_l(\Lambda)\triangleq l(\mathbb{R}^n/\Lambda)\subset\mathbb{R}^n$. Clearly, the set of translates $\{R_l(\Lambda)+b:b\in\Lambda\}$ forms a regular tessellation of $\mathbb{R}^n$, and hence the volume of $R_l(\Lambda)$ equals $V(\Lambda)$, the volume of $n$-space per point of $\Lambda$.

Since $\mathbb{R}^n$ is a normed space under the usual $L^2$-norm, it is common to take elements with minimum norm as coset representatives. A labeling function $V$ such that
\begin{equation}
|V(A)| \leq |a|, \;\forall\, A\in\mathbb{R}^n/\Lambda, a\in A \nonumber
\end{equation}
defines then a fundamental region $R_V(\Lambda)$ known as \emph{fundamental Voronoi region}. The corresponding tessellation of $\mathbb{R}^n$ consists of decision regions for a minimum-distance quantizer (or decoder) that uses $\Lambda$ as codebook.

The role of the labeling function becomes evident when the induced group structure on $R_l(\Lambda)$ is considered. The labeling function is in fact invertible in $R_l(\Lambda)$, and hence, upon defining the sum operation in $R_l(\Lambda)$ as
\begin{equation}
\label{e:Voronoi_sum}
\alpha+\beta \triangleq l\left(l^{-1}(\alpha)+l^{-1}(\beta)\right), \;\forall \alpha,\beta\in R_l(\Lambda)\;,
\end{equation}
$l$ is an isomorphism. Denote with $\nu$ the natural homomorphism $\nu:\mathbb{R}^n \to \mathbb{R}^n/\Lambda$, and define the function $s_l \triangleq l\circ\nu:\mathbb{R}^n \to R_l(\Lambda)$. Once a labeling function $l$ is defined, this homomorphism identifies indeed the coset to which any element of $\mathbb{R}^n$ belongs. As a consequence, $s_l(a)$ represents the \emph{\cvs} of $a\in\mathbb{R}^n$, by analogy with the role of the traditional syndrome in linear codes.

The {\cvs} satisfies the following properties:
\begin{eqnarray}
\label{e:s_periodic}
s_l(a+\lambda) &=& s_l(a), \;\forall a\in\mathbb{R}^n, \lambda\in\Lambda\;;\\
\label{e:s_identity}
s_l(a) &=& a, \;\forall a\in R_l(\Lambda)\;;\\
\label{e:s_idempotent}
s_l\left( s_l(a) \right) &=& s_l(a), \;\forall a\in\mathbb{R}^n\;;\\
\label{e:s_linear}
s_l(a+b) &=& s_l(a)+s_l(b), \;\forall a,b\in\mathbb{R}^n\;.
\end{eqnarray}
In particular, (\ref{e:s_periodic}) and (\ref{e:s_identity}) follow directly from the definition of $s_l$, and state that $s_l$ is \emph{periodic} and that the restriction ${s_l}_{|R_l(\Lambda)}:R_l(\Lambda) \to R_l(\Lambda)$ of $s_l$ is an \emph{identity} respectively; (\ref{e:s_idempotent}) follows directly from (\ref{e:s_identity}), and states that $s_l$ is \emph{idempotent}, i.e.~that $s_l^m=s_l$, $\forall m=1,2,\dots$; (\ref{e:s_linear}), in which the sum on the right-hand side is intended as defined in (\ref{e:Voronoi_sum}), follows from the fact that $s_l$ is a homomorphism. As a remark, $\tilde{s}_l:\mathbb{R}^n \to (\mathbb{R}^n,+)$ is not a homomorphism\footnote{For the sake of clarity, the {\cvs} $s_l(\cdot)$, when intended as belonging to $(\mathbb{R}^n,+)$ rather than $(R_l(\Lambda),+)$, will be hereinafter indicated as $\tilde{s}_l(\cdot)$.} since $R_l(\Lambda)$ is not a subgroup of $(\mathbb{R}^n,+)$.\footnote{For the same reason, while the identity function ${s_l}_{|R_l(\Lambda)}$ is obviously a homomorphism $R_l(\Lambda)\to (R_l(\Lambda),+)$ (in particular, an automorphism), it is not a homomorphism $R_l(\Lambda)\to (\mathbb{R}^n,+)$.} Hence, in general, if both the sums are taken in $(\mathbb{R}^n,+)$, $\tilde{s}_l(a+b)\ne \tilde{s}_l(a)+\tilde{s}_l(b)$. However, it is straightforward to show that
\begin{equation}
\label{e:s_sum}
\tilde{s}_l(a+b)=\tilde{s}_l\left(\tilde{s}_l(a)+\tilde{s}_l(b)\right), \;\forall a,b\in\mathbb{R}^n\;,
\end{equation}
which in turn shows that $\tilde{s}_l(a+b)= \tilde{s}_l(a)+\tilde{s}_l(b)$ if and only if $\tilde{s}_l(a)+\tilde{s}_l(b)\in R_l(\Lambda)$.

In order to evaluate the {\cvs} for a given $a\in\mathbb{R}^n$, it is immediate to verify that the syndrome $s_V(a)\in R_V(\Lambda)$ relative to a fundamental Voronoi region can be obtained as \emph{quantization error} of a minimum-distance quantizer that uses $\Lambda$ as codebook. In particular, defining
\begin{equation}
Q_\Lambda(a) \triangleq \lambda\in\Lambda:|\lambda-a|\leq|\gamma-a|,\;\forall\gamma\in\Lambda \nonumber
\end{equation}
as the closest lattice point to $a$ (with the further condition $(a-\lambda)\in R_V(\Lambda)$ in case of ambiguity), we have
\begin{equation}
\label{e:quantizer_syn}
\tilde{s}_V(a)=a-Q_\Lambda(a)\;.
\end{equation}
In this case, as the traditional syndrome, the {\cvs} of $a$ identifies the minimum-norm element that, subtracted to $a$, leads to an element (the closest) of $\Lambda$.

\section{{\DSC} and Channel Coding}\label{s:dsc_review}
After a brief review of the well known concepts of \emph{\dsc} and \emph{\csi}, this section discusses how these problems are intertwined with the more traditional channel coding problem. In particular, this fact is due to the existence of a virtual \emph{correlation} channel, which can be seen both as \emph{forward} or \emph{backward} channel.

\subsection{{\DSC} and \CSI}
\begin{figure}%
\centering
\includegraphics[scale=0.8]{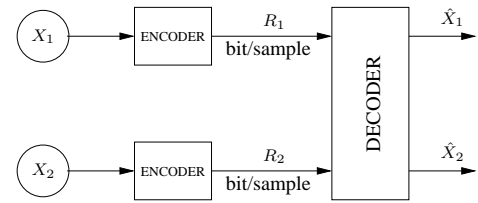}
\caption{Distributed coding of two random sources.}
\label{f:dsc}
\end{figure}

The notion of \emph{\dsc} refers to the problem of coding the outcomes of a random vector $\mathbf{X}=\left[X_i\right]_{i=1,2,\dots,N}$ by using $N$ independent encoders that cannot collaborate with each other, but whose respective $N$ outputs are jointly fed to a single decoder. Assume that the random variables $X_i$ have a discrete alphabet. Then, by the \emph{source coding theorem} \cite{cover_ElemInfoThNew}, perfect reconstruction is asymptotically achieved if each encoder can communicate with a rate greater or equal than the entropy rate $H_\infty(X_i)$ of $X_i$. However, because of joint decoding, this is no longer a necessary condition. Slepian and Wolf showed that in case of independent and identically distributed outcomes with $N=2$ (see Fig.~\ref{f:dsc}), the necessary (and sufficient) conditions are \cite{slepian_noiselessCodingCorrSrc}
\begin{eqnarray}
R_1 &\geq& H(X_1|X_2)\;, \nonumber\\
R_2 &\geq& H(X_2|X_1)\;, \nonumber\\
R_1+R_2 &\geq& H(X_1,X_2)\;, \nonumber
\end{eqnarray}
where $H(\cdot|\cdot)$ and $H(\cdot,\cdot)$ denote the conditional and the joint entropy respectively.

\begin{figure}%
\centering
\includegraphics[scale=0.8]{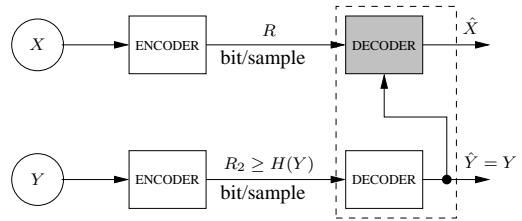}
\caption{Coding with side information at the decoder.}
\label{f:csi}
\end{figure}

In such a case, denote $X_1$, $X_2$, and $R_1$ with $X$, $Y$, and $R$ respectively. If $R_2\geq H(Y)$, i.e.~if the decoder can perfectly reconstruct the outcomes of $Y$ upon receiving the output of the second encoder (see Fig.~\ref{f:csi}), the first encoder just needs to communicate with a rate equal to $H(X|Y)\leq H(X)$ for perfect reconstruction, since the shaded decoder in Fig.~\ref{f:csi} can rely on the \emph{side information} $Y$ correlated with $X$. Given a certain upper bound $D$ to the distortion between $\hat{X}$ and $X$ (that can be discrete or continuous random variables), the general problem of finding the minimum achievable value for $R$ goes under the name of \emph{\csi} at the decoder. This problem was solved by Wyner and Ziv \cite{wyner_RDFunctSrcCodingSideInfo}, and it turned out that the obtained rate-distortion function $R_{X|Y}^\ast(D)$ is in general greater or equal than the rate-distortion function $R_{X|Y}(D)$ corresponding to the case where $Y$ is as well accessible by the encoder.

\subsection{Connections to Channel Coding}\label{s:connection_review}
Since the ultimate performance limits shown in \cite{slepian_noiselessCodingCorrSrc,wyner_RDFunctSrcCodingSideInfo} were obtained with non-constructive proofs, effective, practical solutions have not appeared until recently. Current approaches, that are getting closer and closer to the theoretical limits \cite{xiong_DSCSensors}, have stemmed from the connection that {\dsc} has to channel coding \cite{wyner_resultsShannon}.

In fact, the statistical dependence between $X$ and $Y$ can be exactly characterized in terms of a virtual \emph{correlation} channel. This channel can be defined in two equivalent ways, as follows.
\begin{enumerate}
  \item \emph{Forward channel} (FCH): the complete statistical description of the tuple $(X,Y)$ is given by the probability mass function\footnote{If the random variables are not discrete, the probability mass functions $p(\cdot)$ can be replaced with probability density functions (\emph{pdf}) $f(\cdot)$.} (\emph{pmf}) $p(x)$ and by the conditional \emph{pmf} $p(y|x)$. Hence, $Y$ can be seen as the \emph{output} of the stochastic channel described by $p(y|x)$ into which $X$, distributed according to $p(x)$, is fed as \emph{input}.
  \item \emph{Backward channel} (BCH): alternatively, the complete statistical description of $(X,Y)$ can be given by the \emph{pmf} $p(y)$ and by the conditional \emph{pmf} $p(x|y)$. In this characterization, $X$ is seen as the \emph{output} of the stochastic channel described by $p(x|y)$ whose \emph{input} is $Y$, distributed according to $p(y)$.
\end{enumerate}

In literature the FCH interpretation is commonly given. In the {\csi} problem, according to this interpretation, a noisy \emph{observation} $Y$ of the \emph{unknown} $X$ is given to the decoder together with some \emph{prior} information about $X$, provided by the encoder. Being aware of the FCH statistics, the encoder must hence provide to the decoder the smallest number of independent \emph{constraints} which allow for reconstruction within the desired distortion.

\begin{figure}%
\centering
\includegraphics[scale=0.8]{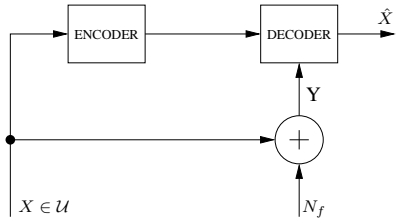}
\caption{Forward additive channel ($N_f$ is independent of $X$).}
\label{f:ddsc_fw}
\end{figure}

In practice, in the discrete case the alphabet of both $X$ and $Y$ coincide with a Galois field $\mathbb{F}$, and the FCH can be described as \emph{additive} channel. This implies that $Y=X+N_f$, with $N_f$ a random variable (\emph{noise}) \emph{independent} of $X$ and distributed according to some \emph{pmf} $p(n_f)$ (see Fig.~\ref{f:ddsc_fw})\footnote{It is worth to remark that this characterization defines a \emph{symmetric} channel in the sense given in \cite{cover_ElemInfoThNew}. Consequently, $H(Y)\geq H(X)$, where, unless $N_f$ is pseudo-aleatory, the equal sign holds if and only if $X$ is uniformly distributed (this condition is necessary in order to achieve the capacity of any symmetric channel and implies that $Y$ is uniformly distributed as well). The additive channel, however, captures neither all possible forms of correlation nor even all possible symmetric correlation channels.}. The connection to channel coding then comes from the fact that there may exist a \emph{zero-error} $(n,k)$-linear code which achieves the capacity $C = \log_2|\mathbb{F}| - H(N_f)$ (bits per channel use) of the FCH, i.e.~such that $k/n\cdot\log_2|\mathbb{F}| = C$. In this case $X$ can be perfectly reconstructed once the coset to which each successive $n$-tuple of outcomes of $X$ belongs is signalled to the decoder. This is simply accomplished by decoding $Y$ into the signalled coset. If $X$ is uniformly distributed on $\mathbb{F}$ this communication requires a rate $R = (n-k)/n\cdot\log_2|\mathbb{F}|$ (bits per channel use), i.e.~it achieves the Slepian-Wolf limit $H(X|Y) = H(Y|X) = H(N_f) = \log_2|\mathbb{F}| - C$.

As discussed in Section \ref{s:syndrome_review}, the \emph{syndrome} $s_H(x)$ can be used to inform the decoder about the coset to which the \emph{unknown} belongs. Details and experimental results regarding this distributed coding strategy called \emph{distributed source coding using syndromes} (DISCUS) can be found in \cite{xiong_DSCSensors,pradhan_DISCUS1,pradhan_DISCUS2,pradhan_genCosetCodesBinning}.

Once the decoder has the side information, the remaining information needed for perfect reconstruction should actually regard the \emph{noise} only. The fact that in the previous approach the communication rate equals exactly $R = H(N_f)$ seems to confirm this observation. However, within the additive FCH interpretation there is no apparent relation between the \emph{noise} and the output of the encoder, which is indeed driven by a signal perfectly \emph{independent} of the \emph{noise} itself.

According to the BCH interpretation, instead, the input of the encoder can be seen as a direct \emph{observation} of the \emph{unknown} noise introduced by the channel, but measured with some \emph{offset} (that is input into the BCH) known at the decoder only. The encoder, hence, given the BCH statistics, should provide the decoder with the smallest amount of information that allows for reconstruction of the \emph{unknown} (i.e.~of the \emph{noise}) within the desired distortion. Obviously, it should operate in a way such that any allowable \emph{offset} does not increase the amount of information needed.

\begin{figure}%
\centering
\includegraphics[scale=0.8]{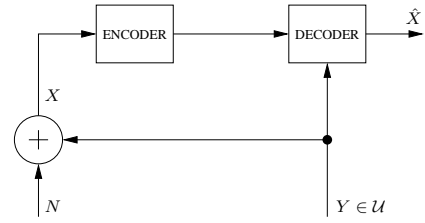}
\caption{Backward additive channel ($N$ is independent of $Y$) equivalent to the forward channel of Fig.~\ref{f:ddsc_fw}.}
\label{f:ddsc_bk}
\end{figure}

Again, in the discrete case the BCH can be usually described as \emph{additive} channel over the finite field $\mathbb{F}$. This implies that $X=Y+N$ (the term \emph{offset} used for the side information becomes now clearer), with $N$ a \emph{noise} \emph{independent} of $Y$ and distributed according to some \emph{pmf} $p(n)$ (see Fig.~\ref{f:ddsc_bk}). If there exist a \emph{zero-error} $(n,k)$-linear code which achieves the capacity $C = \log_2|\mathbb{F}| - H(N)$ (bits per channel use) of the BCH (i.e.~such that $k/n\cdot\log_2|\mathbb{F}| = C$), $X$ can be perfectly reconstructed once the coset to which each successive $n$-tuple of outcomes of $X$ belongs is signalled to the decoder. By knowing the \emph{offset}, in fact, the decoder easily computes first the coset to which the \emph{noise} belongs, and then the \emph{noise} itself. Since this communication requires a rate up to $R = (n-k)/n\cdot\log_2|\mathbb{F}|$ (bits per channel use), the Slepian-Wolf limit $H(X|Y) = H(N) = \log_2|\mathbb{F}| - C$ is exactly achieved.

Apparently, in the discrete case examined above, if both the additive FCH and BCH interpretations are applicable it is not yet clear why one should prefer one over the other. In this paper it is claimed that the BCH interpretation is preferable because of the following.
\begin{itemize}
  \item The existence of a good linear code for the additive FCH does not guarantee the achievability of the Slepian-Wolf limit when $X$ is not uniformly distributed. For example, it is always possible to assign non-uniform probabilities $p(x)$ in a way such that each coset is still equiprobable. Consequently, the transmission rate needed will still equal $H(N_f)$, even if in this case $H(Y)>H(X)$ implies $H(X|Y)=X(Y|X)+H(X)-H(Y)<H(Y|X)=H(N_f)$.
  \item If a good linear code for the additive BCH exists, then the Slepian-Wolf limit $H(X|Y)=H(N)$ is always achieved, independently of the \emph{pmf} of $Y$. Moreover, during the design of the distributed coding system, it is not necessary to know $p(y)$, but only $p(n)$, i.e.~the BCH statistics.
  \item Finding a good linear code for the additive BCH is equivalent to looking for a good linear encoder for $N$ (i.e.~a good linear encoder for $X$ when $Y=0$). In fact, the latter search implies to find a sufficiently large $n$ such that $n$-tuples of $N$ distribute uniformly into the \emph{typical set} $\mathcal{A}^{(n)}\subset\mathbb{F}^n$, and a linear function $s:\mathbb{F}^n\to\mathbb{F}^{n-k}$ whose restriction to $\mathcal{A}^{(n)}$ is one-to-one into $\mathbb{F}^{n-k}$ (and hence with $k$ such that $|\mathbb{F}|^{n-k}=|\mathcal{A}^{(n)}|\simeq 2^{nH(N)}$). The dual code of $s(\mathbb{F}^n)$ is then a good linear code for the additive BCH.
\end{itemize}
Within the BCH interpretation, the encoder design is hence directly tailored to the \emph{noise} $N$. Then, since encoding is a linear operation, any \emph{offset} known by the decoder will not harm the performance of the system. In fact, the decoder can easily remove its effect on the information received by the encoder, which is still uniformly distributed on $\mathbb{F}^{n-k}$, and hence requires the same transmission rate.

It is worth to point out that in the discrete and finite case both forward and backward interpretations give additive channels (with input-noise independence) if and only if $X$ and $Y$ are uniformly distributed, unless $N_f$ (or $N$) is pseudo-aleatory (in which case the channel is deterministic and one-to-one). In particular, if an additive FCH exists and $X$ is uniformly distributed, then $N=-N_f$ is independent of $Y$, which in turn is uniformly distributed. Viceversa, if an additive BCH exists and $Y$ is uniformly distributed, then $N_f=-N$ is independent of $X$, which in turn is uniformly distributed. In practice, if $N=-N_f$, the two schemes in Fig.~\ref{f:ddsc_fw} and Fig.~\ref{f:ddsc_bk} are equivalent. The point here is that any good linear code for the additive FCH turns out to be equally good for the BCH. In other words, the DISCUS system that is tailored for an additive FCH with a uniformly distributed source is also optimal for the corresponding BCH with any desired side information distribution. Even if in the discrete case it is not necessary to explicitly design codes using the BCH interpretation, this approach will turn out to be more useful for the continuous channels discussed in the following.

\subsection{Continuous Correlation Channels}\label{s:continue_chn}
Assuming that $X$ and $Y$ are continuous random variables that take values on the Euclidean space $\mathbb{R}$, the virtual channel (FCH or BCH) that describes the mutual correlation is a continuous channel. In this case, in general, as $D\to 0^+$, $R^\ast_{X|Y}(D)\geq R_{X|Y}(D)\to +\infty$, i.e.~in the problem of {\csi} perfect reconstruction is practically not achievable, unless we allow the encoder to communicate at an infinite rate. In the following, assuming for a while that this is possible, the generalization to this domain of the results of the previous sub-section is discussed\footnote{This generalization is not straightforward. The differences do not essentially follow from the fact that the domain is \emph{continuous}, but rather arise because the space $\mathbb{R}$ has a \emph{non-finite} measure. In this case, the uniform distribution cannot be defined \emph{on the entire alphabet}, and hence the capacity of the additive channels cannot be achieved by such a \emph{pdf}. Consequently, the \emph{pdf} of the input and of the output of the channel when the capacity is achieved differ. Instead, there exist additive channels in a continuous but \emph{finite}-measure domain (as for example the mod-$\Lambda$ channel defined in \cite{forney_SphereBoundAchievingCosetCodes}) where the duality with the discrete and finite case would be more strict.}.

\begin{figure}%
\centering
\includegraphics[scale=0.8]{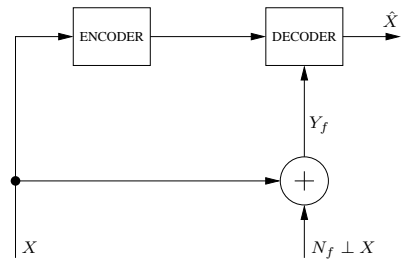}
\caption{Forward continuous additive channel ($N_f$ is uncorrelated with $X$).}
\label{f:dsc_fw}
\end{figure}

In particular, assume that the FCH is actually an additive channel such that $Y_f=X+N_f$, with $N_f$ independent of $X$ (see Fig.~\ref{f:dsc_fw})\footnote{It is hereinafter assumed that the channel is not deterministic and that, without loss of generality, the continuous random variables have zero mean.}. Suppose that a lattice $\Lambda$ exists in $\mathbb{R}^n$ whose intersection with a bounded set $\mathcal{A}\subset\mathbb{R}^n$ defines a capacity-achieving code $\Lambda_b$ (subject to some cost constraint) for the FCH. The volume of $n$-space per point of $\Lambda$ approximates the volume of the typical set of $N_f$, because in \emph{each} coset of $\mathbb{R}^n/\Lambda$ there is a \emph{unique noise} representative. Perfect reconstruction (with probability close to one) is achieved by channel-decoding $Y_f$ into the coset signalled to the decoder as the coset to which any $n$-tuple of outcomes of $X$ belongs. It would be for example possible to reconstruct $X$ with negligible probability of error by transmitting a {\cvs} $s_l(x)$ as defined in Section \ref{s:c_syndrome_review}.

However, differently from the discrete case, $Y_f$ and $N_f$ are neither independent nor uncorrelated ($E[Y_fN_f]=E[N_f^2]\ne 0$), not even when the \emph{pdf} of $X$ achieves channel capacity. Hence, the information brought by $Y_f$ about the \emph{noise} may relax the need for transmission of as many different messages as the possible realizations of $N_f$. In the example of an additive Gaussian channel with $N_f\in\mathcal{N}(0,\sigma_f^2)$, the volume of $n$-space per point of $\Lambda$ is approximately $V(\Lambda)\simeq 2^{nh(N_f)}=2^{nh_G(\sigma_f^2)}$, where $h(\cdot)$ and $h_G(\cdot)$ denote the differential entropy and the differential entropy of the Gaussian random variable (i.e.~the upper bound of the differential entropy under the specified power constraint) respectively. Invoking a sphere-packing argument, if $X\in\mathcal{N}(0,\sigma_x^2)$ (which achieves the capacity under the constraint $E[X^2]\leq\sigma_x^2$) $Y_f$ is used at the decoder to discriminate, on average, between about $2^{nh_G(\sigma_x^2)}/V(\Lambda)\simeq 2^{\frac{n}{2}\log_2(\sigma_x^2/\sigma_f^2)}$ different $n$-tuples. But actually the number of distinct messages per $n$ channel uses that could be reliably transmitted through the FCH under the same power constraint is $2^{nC}=2^{\frac{n}{2}\log_2(1+\sigma_x^2/\sigma_f^2)}>2^{\frac{n}{2}\log_2(\sigma_x^2/\sigma_f^2)}$.

In other words, there must exist a denser code $\Lambda'_b$ (which is assumed to be a bounded subset of a lattice $\Lambda'\subset\mathbb{R}^n$) such that the relative coset information is still sufficient for perfect reconstruction. Since at least a coset of $\mathbb{R}^n/\Lambda'$ contains \emph{more than one} representative of $N_f$, for this code the algorithm sketched above (channel-decoding of $Y_f$ into the coset signalled by the encoder) no longer leads to perfect reconstruction.

\begin{figure}%
\centering
\includegraphics[scale=0.8]{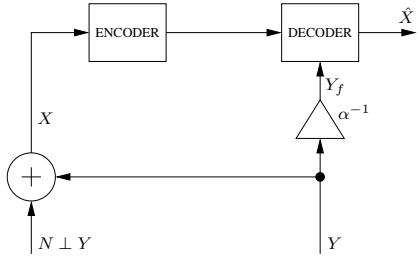}
\caption{Backward continuous additive channel ($N$ is uncorrelated with $Y$) equivalent to the forward channel of Fig.~\ref{f:dsc_fw}.}
\label{f:dsc_bk}
\end{figure}

As noted in the previous section, a slightly different perspective is possible when the correlation is described in terms of a BCH. In the continuous domain, under the condition that $X$ and $N_f$ are uncorrelated, any additive FCH can be interpreted as additive BCH as follows. Define
\begin{equation}
\label{e:bk_Y}
Y\triangleq\alpha Y_f\;,
\end{equation}
with
\begin{equation}
\alpha=\frac{E[X^2]}{E[X^2]+E[N_f^2]}<1\;, \nonumber
\end{equation}
as the optimum linear predictor of $X$ from $Y_f$. Consequently,
\begin{equation}
N\triangleq X-Y=X-\alpha Y_f=(1-\alpha)X-\alpha N_f \nonumber
\end{equation}
is uncorrelated with $Y$, and $X=Y+N$. Hence, if $Y_f$ is multiplied by $\alpha$, then the \emph{forward} channel of Fig.~\ref{f:dsc_fw} can be interpreted as a \emph{backward} channel (see Fig.~\ref{f:dsc_bk}), with \emph{input}, \emph{noise} and \emph{output} variances equal to
\begin{eqnarray}
E[Y^2] &=& \alpha^2 E[Y_f^2]=\alpha E[X^2] \label{e:input_ch}\\
E[N^2] &=& E[X^2]-E[Y^2]=(1-\alpha)E[X^2]=\alpha E[N_f^2]\;, \label{e:noise_ch}\\
E[X^2] &=& \alpha(E[X^2]+E[N_f^2])=\alpha E[Y_f^2]\;. \nonumber
\end{eqnarray}
Moreover, if $X$ and $N_f$ are independent and $X$ is distributed according to the \emph{pdf} that achieves the FCH capacity, then it is reasonable for $N$ to be independent of $Y$. This is certainly true for additive Gaussian channels.

Using this interpretation, and assuming that the \emph{pdf} of $X$ is the one that achieves the FCH capacity, a code $\Lambda'_b$ is found such that signalling the coset of $\mathbb{R}^n/\Lambda'$ to which any $n$-tuple of outcomes of $X$ belongs is sufficient for perfect reconstruction at the decoder. In the Gaussian case with $N_f\in\mathcal{N}(0,\sigma_f^2)$, for example, the scaled code $\Lambda'_b=\alpha^{1/2}\Lambda_b$ can be used. Observe that it represents a capacity-achieving code for the BCH with the power constraint $E[Y^2]\leq\alpha\sigma_x^2$ (that $Y$ as defined in (\ref{e:bk_Y}) satisfies), which has the same capacity $C=\frac{1}{2}\log_2(1+\sigma_x^2/\sigma_f^2)$ of the FCH with the power constraint $E[X^2]\leq\sigma_x^2$. Hence, with probability close to one, any other $n$-dimensional realization of the \emph{noise} $N$ belongs to a different coset of $\mathbb{R}^n/\Lambda'$, such that in \emph{each} coset there is a \emph{unique} \emph{noise} representative. Being independent of $N$, $Y$ ($Y_f$) does not give any other useful information about $N$. As in the discrete case, since the information about the \emph{noise} can be \emph{linearly} formed at the encoder with an \emph{offset} known at the decoder, perfect reconstruction is achieved. In particular, this is accomplished by channel-decoding $Y$ (in place of $Y_f$) at the decoder. Now that the volume of $n$-space per point of $\Lambda'$ is approximately $V(\Lambda')\simeq 2^{nh(N)}=2^{nh_G(\alpha\sigma_f^2)}$, it turns out that the number of distinct messages per $n$ channel uses that could be reliably transmitted through the correlation channel is correctly $2^{nh_G(\sigma_x^2)}/V(\Lambda')\simeq 2^{\frac{n}{2}\log_2(\alpha^{-1}\sigma_x^2/\sigma_f^2)}=2^{nC}$.

While in the discrete and finite case any good linear code for the additive FCH turns out to be equally good for the BCH, in the continuous case the coset information sent to the decoder should be relative to a code, tailored to the BCH statistics, which is denser than the former. This approach will be discussed in the next section, where it will be showed that if a good channel code is available for the additive Gaussian BCH, then the rate-distortion function $R^\ast_{X|Y}(D)$ can be approximately achieved by sending a \emph{quantized} version of the {\cvs}, independently of the \emph{pdf} of $Y$.

\section{{\CVS}-Based {\DSC}}\label{s:algorithm}
In the last fifteen years, the binary codes that have been introduced in the literature, such as the turbo-codes \cite{berrou_turboJCOM} and the low-density parity-check (LDPC) codes \cite{mackay_ldpc}, come very close to the channel capacity of the binary additive channel with a reasonable computational complexity. But, in many {\dsc} applications, $X$ and $Y$ take binary values, and their correlation can be analogously described by a virtual additive channel. Hence, since soon after the discovering of the strong connection to channel coding discussed in Section \ref{s:connection_review}, several approaches have been successfully proposed for the problem of {\dsc} based on these linear channel codes \cite{aaron_compressionSideTurboCodes,liveris_sideLDPC}.

Following the success of these coding algorithms in the binary domain, if continuous (or discrete, but non-binary) random variables have to be coded, a transformation into the binary domain is usually applied prior to the actual coding operation, for example by first quantizing the variables and by then scanning their values by bit-planes \cite{girod_DVC}.

\begin{figure}%
\centering
\includegraphics[scale=0.8]{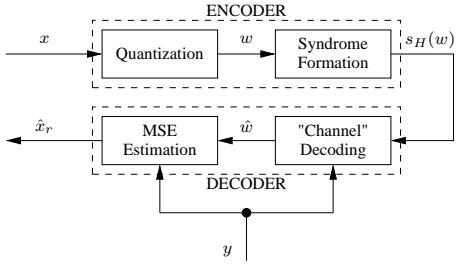}
\caption{Coding with side information at the decoder, according to the DISCUS system.}
\label{f:syndrome}
\end{figure}

Otherwise, it is possible to consider non-binary discrete linear codes, such as the trellis codes (\cite{forney_CosetI,forney_CosetII}), to directly code discrete random variables or quantized versions of continuous random variables. For example, the DISCUS system \cite{pradhan_DISCUS2} is actually concerned with the problem of {\csi} a continuous variable $X$, which is first quantized into a discrete domain, in which cosets of some trellis code \cite{marcellin_TCQ} are identified. For any $n$-dimensional realization $x$ of $X$, the encoder sends the syndrome $s_H(w)$ that identifies the coset to which the quantized version $w$ of $x$ belongs (see Fig.~\ref{f:syndrome}). Since the version $\hat{w}$ of $w$ reconstructed at the decoder lies in the discrete domain, then a minimum square-error (MSE) estimate $\hat{x}_r$ is found, using as well the same side information $y$ which by channel-decoding led to $\hat{w}$.

While these coding systems achieve somewhat a good performance, the interplay between the quantizer and the syndrome former \cite{pradhan_DISCUS2,rebollo_optQuantDSC} does not usually lead to a straightforward optimal design procedure once the correlation and the available transmission rate are known. According to the discussion in Section \ref{s:continue_chn}, by using a {\cvs} it is instead possible to form a syndrome which directly depends on the continuous realization to be coded. If the syndrome formation is exactly tailored to the BCH correlation, then upon receiving that information perfect reconstruction is possible. Otherwise, if that information cannot be entirely sent due to some transmission rate constrain, it will be still possible to send a quantized version of it. The difference here is that the design of the system exactly depends on the knowledge of the correlation (responsible for the syndrome formation) and on the knowledge of the available transmission rate (responsible for the syndrome quantization).

\begin{figure}%
\centering
\includegraphics[scale=0.8]{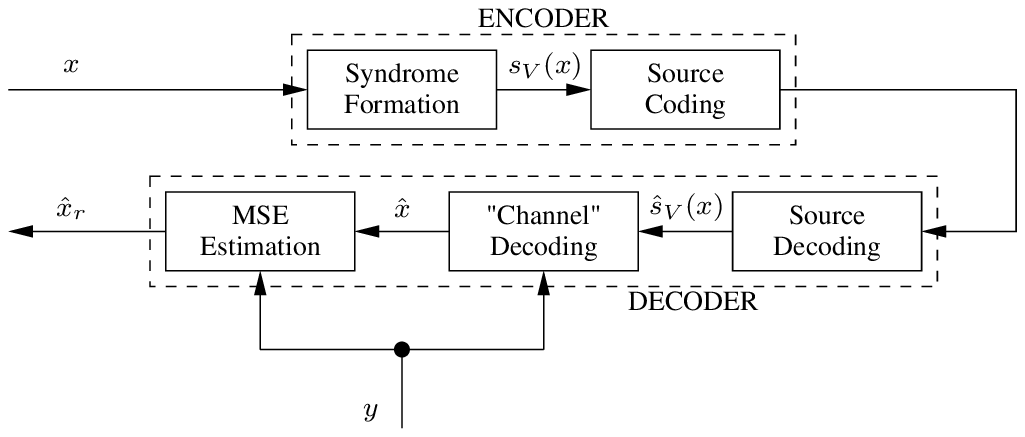}
\caption{Coding with side information at the decoder using {\cvs}s.}
\label{f:c_syndrome}
\end{figure}

The proposed coding system is sketched in Fig.~\ref{f:c_syndrome}. The {\cvs} $s_V(x)$ is formed in correspondence of any $n$-dimensional realization $x$ of $X$, and an approximation $\hat{s}_V(x)$ is received at the decoder, which estimates $\hat{x}$ by channel-decoding of $y$ (that represents the input of the BCH) and finally reconstructs $\hat{x}_r$ as a MSE estimate of $x$ from $\hat{x}$ and $y$.

In the case that a capacity achieving linear code (i.e.~a bounded subset $\Lambda_b$ of a lattice $\Lambda$) exists for the BCH and that the corresponding {\cvs} $s_V(x)$ is correctly received at the decoder, perfect reconstruction of $x$ is asymptotically achieved when $n\to\infty$. This is shown by the following expressions
\begin{eqnarray}
x &=& y + n \nonumber\\
  &=& y + \tilde{s}_V(n) \label{e:n_in_cell}\\
  &=& y + \tilde{s}_V(x - y) \nonumber\\
  &=& y + \tilde{s}_V\left( \tilde{s}_V(x) + \tilde{s}_V(-y) \right) \label{e:i_sum}\;,
\end{eqnarray}
where (\ref{e:n_in_cell}) follows from (\ref{e:s_identity}) and from the fact that all \emph{noise} realizations tend to lie uniformly in the fundamental Voronoi region of $\Lambda$, and (\ref{e:i_sum}) follows from the property of the sum (\ref{e:s_sum}). Even if equation (\ref{e:i_sum}) already shows the sufficiency of $s_V(x)$ for perfect reconstruction, from an operative point of view it is useful to note that
\begin{eqnarray}
x &=& y + \tilde{s}_V\left( \tilde{s}_V(x) - y - Q_\Lambda(-y) \right) \label{e:i_qs}\\
  &=& y + \tilde{s}_V\left( \tilde{s}_V(x) - y \right) \label{e:i_periodic} \\
  &=& \tilde{s}_V(x) - Q_\Lambda\left( \tilde{s}_V(x) - y \right) \label{e:i_qsa}\;,
\end{eqnarray}
where (\ref{e:i_qs}) follows from the operative definition of the {\cvs} (\ref{e:quantizer_syn}), (\ref{e:i_periodic}) follows from the periodicity (\ref{e:s_periodic}) of $s_V(\cdot)$, and (\ref{e:i_qsa}), again, from the definition (\ref{e:quantizer_syn}). Then, a single quantization operation is sufficient at the decoder for perfect reconstruction.

Assuming now that the channel-decoding algorithm is the same even if the received information from the encoder consist of a quantized (\emph{noisy}) version of the {\cvs}, at the decoder the following hold
\begin{eqnarray}
\hat{x} &=& y + \tilde{s}_V\left( \tilde{s}_V(x) + q + \tilde{s}_V(-y) \right) \label{e:noisy_s}\\
        &=& y + \tilde{s}_V\left( \tilde{s}_V(x) + \tilde{s}_V(q) + \tilde{s}_V(-y) \right) \label{e:q_in_cell}\\
        &=& y + \tilde{s}_V( x + q - y ) \label{e:i_suma} \\
        &=& y + \tilde{s}_V( n + q ) \nonumber\\
        &=& x + \left( q - Q_\Lambda( n + q )\right) = x + (q + q_{ol})\label{e:x_r} \;,
\end{eqnarray}
where: (\ref{e:noisy_s}) is obtained from (\ref{e:i_sum}) by substitution of $s_V(x)$ with $\hat{s}_V(x)=\tilde{s}_V(x) + q$ ($q$ is an $n$-dimensional realization of a random variable $Q$); (\ref{e:q_in_cell}) follows from (\ref{e:s_identity}) and from the reasonable assumption that all the realizations of $q$ lie in same region as $n$; (\ref{e:i_suma}) and (\ref{e:x_r}) follow, again, from the sum-property (\ref{e:s_sum}) and from the definition (\ref{e:quantizer_syn}) respectively.

The total reconstruction error $q_t\triangleq q+q_{ol}$ is hence the sum of the \emph{granular} error $q$, and of the \emph{overload} error $q_{ol}\triangleq - Q_\Lambda(n + q)$. If $q$ is negligible w.r.t.~$n$, and it is reasonable that this happens at high transmission rates (recall that if $\Lambda_b$ is capacity-achieving $s_V(X)$ is distributed as $N$), then the probability of $q_{ol}$ being not zero (that is called \emph{error probability}\footnote{An error occurs each time some coordinate of the $n$-dimensional vector $Q_\Lambda(n+q)$ is not $0$.} $P_e$) is negligible and $q_t\simeq q$. Hence, the same additive \emph{error} $Q$ would impair both the syndrome and the source at the decoder (as it was shown in a different way in \cite{cappellari_DSCcontinueSyndrome}). This in turn shows that the operational rate-distortion function obtained by the proposed coding system $\hat{R}^\ast_{X|Y}(D)$ satisfies at high rates
\begin{equation}
\hat{R}^\ast_{X|Y}(D)\simeq \hat{R}_{S_V}(D)\; \nonumber,
\end{equation}
where $\hat{R}_{S_V}(D)$ is the operational rate-distortion function relative to the source encoder used to code the {\cvs}. Again, since $s_V(X)$ is distributed as $N$, the ultimate performance achievable is then represented by $R_N(D)$, the rate-distortion function relative to the \emph{noise} added by the BCH.

This is particularly interesting when the \emph{noise} $N$ (which is the \emph{difference of} $X$ and $Y$) is Gaussian (and of course independent of $Y$), i.e.~$N\in\mathcal{N}(0,\sigma_n^2)$. In this case, as shown first in \cite{pradhan_dualitySrcChnCodSideInfo} and then in \cite{cheng_SuccessiveRefWynerZiv}, $R^\ast_{X|Y}(D)=R_N(D)$ independently of the distribution of $Y$ (in particular, without requiring $Y$ to be Gaussian as first assumed in \cite{wyner_RDFunctSrcCodingSideInfo}). Consequently,
\begin{equation}
\hat{R}^\ast_{X|Y}(D)\geq R_N(D)=R^\ast_{X|Y}(D)\; \nonumber,
\end{equation}
i.e.~in the Gaussian case and at high rates, \emph{the ultimate performance of the proposed coding algorithm is the rate-distortion function with side information at the decoder itself}. This result is achieved upon utilizing a capacity-achieving \emph{channel} code $\Lambda_b$ (i.e.~upon utilizing a rate-distortion optimal \emph{source} code \cite{pradhan_dualitySrcChnCodSideInfo}) for syndrome formation, and a rate-distortion optimal \emph{source} coder for syndrome encoding.

In case the rate is not sufficiently high for this to hold, $P_e\ne 0$ and an MSE estimate can reduce the global error. In particular, observing that $X=Y+N$ (with $N$ independent of $Y$) and $\hat{X}= X+Q_t$ (with $Q_t$ assumed independent of both $Y$ and $N$ and with variance $\sigma_{q_t}^2$), it is perfectly correct to use
\begin{equation}
\hat{X}_r = \frac{\sigma_n^2}{\sigma_n^2+\sigma_{q_t}^2}\hat{X} + \frac{\sigma_{q_t}^2}{\sigma_n^2+\sigma_{q_t}^2}Y\label{e:MSE_estimate}
\end{equation}
as optimal linear prediction of $X$ from $\hat{X}$ and $Y$ at the decoder; the residual error variance $\sigma_r^2$ would satisfy then $1/\sigma_r^2 = 1/\sigma_{q_t}^2 + 1/\sigma_n^2$ \cite{gersho_VQ}.

The existence of effective source coders for Gaussian random variables, such as the {\tcqr}s \cite{marcellin_TCQ}, suggests the feasibility of their application in this framework, in the syndrome encoder as well as in the channel decoder. This application, with the relative experiments, is discussed in the next section.

\section{Experiments}\label{s:experiments}
In this section, the problem of coding a source when another one is available only at the decoder is targeted with the approach outlined in Section \ref{s:algorithm}. In particular, it is assumed that the correlation is modeled by a \emph{backward} Gaussian additive channel (Fig.~\ref{f:dsc_bk}) with $N\in\mathcal{N}(0,\sigma_n^2)$. The variance $\sigma_n^2$ of the Gaussian \emph{noise} to be generated is chosen in order to guarantee a given \emph{backward correlation signal-to-noise ratio} (BC-SNR), i.e.~a certain ratio between the variance $\sigma_y^2$ of the side information $Y$ and $\sigma_n^2$. In the experiments, $\sigma_y^2$ is fixed (and equals $1$), but both the Gaussian and the uniform distribution has been tested for $Y$, obtaining as expected exactly the same results.

As a remark, note that if $Y\in\mathcal{N}(0,\sigma_y^2)$, then $X\in\mathcal{N}(0,\sigma_x^2)$, with $\sigma_x^2=\sigma_y^2+\sigma_n^2$, and the simulated additive BCH can be assumed as derived from a Gaussian additive FCH with Gaussian input (as discussed in Section \ref{s:continue_chn}), which is the problem treated in \cite{pradhan_DISCUS2}. For comparison purposes, observe that the \emph{correlation signal-to-noise ratio} (C-SNR) of the corresponding FCH equals the BC-SNR of the simulated BCH since $\sigma_x^2/\sigma_f^2=\sigma_y^2/\sigma_n^2$, as given by (\ref{e:input_ch}) and (\ref{e:noise_ch})\footnote{The only difference between the two approaches is then in the fact that $\sigma_x^2$ is kept constant in \cite{pradhan_DISCUS2}, while it is $\sigma_y^2$ that is kept constant in this paper.}.

\subsection{{\CVS} Formation}
As \emph{\tcq} (TCQ) based on the partition $a\mathbb{Z}/4a\mathbb{Z}$ \cite{marcellin_TCQ} defines a \emph{geometrically uniform} source code \cite{forney_GeomUnifCodes}, and in particular a lattice $\Lambda$ \cite{forney_CosetI}, the encoder computes the {\cvs} of the source as quantization error after TCQ, as given in (\ref{e:quantizer_syn}). The number of samples jointly processed is $n=1000$, and the tested codes are the ones in \cite{marcellin_TCQ} relative to $8$-, $64$-, and $256$-state trellises.

The experimental second moment per dimension of the Voronoi region relative to the TCQ codes measured in \cite{marcellin_TCQ} is used to determine the \emph{scaling factor} $a$ in a way such that the {\cvs} variance equals the \emph{noise} variance $\sigma_n^2$ multiplied by a \emph{volumetric factor} $K$. This factor takes into account for the non-optimality of the TCQ codes for channel coding and it is numerically estimated. For a capacity-achieving code, $s_V(X)$ would be distributed as $N$ and hence have the same variance ($K=1$); in practice, using sub-optimal codes, the normalized volume (per two dimensions) of the non-spherical Voronoi region must be larger ($K>1$) for optimal performance.

\subsection{{\CVS} Coding}
Any suitable source coding algorithm can be used to code the $n$-dimensional {\cvs} $s_V(x)$ into its approximation $\hat{s}_V(x)$ at the desired transmission rate. In this paper, since the focus is on practical applications, the target rate $R$ is chosen in the range $0.5\div 3.0$ bit/sample. Again, TCQ has been preferred over other coding systems because of its simplicity and in order to reutilize the same tool used in syndrome formation. In particular, independently of the TCQ system used to obtain $s_V(x)$, two TCQ systems have been tested, both based on $8$-state trellises.

\begin{enumerate}
  \item $\mathbb{Z}/4\mathbb{Z}$-based TCQ. Given the target rate $R$, a regular TCQ based on the partition $b\mathbb{Z}/4b\mathbb{Z}$ is designed (by dimensioning $b$) such that its normalized volume is $2^{2R}$ times less than the normalized volume relative to the TCQ system used in syndrome formation.
  \item Rate-distortion optimized TCQ. Once $R$ is assigned, a finite-alphabet TCQ system achieving the rate $\lceil R\rceil$ (that uses $2^{\lceil R\rceil+1}$ reconstruction points \cite{marcellin_TCQ}) is iteratively optimized for a set of $100$ syndromes that serve as \emph{training sequences}. The used algorithm is an adaptation to the trellis structure of the algorithm presented in \cite{chou_ECVQ}. In particular, all reconstruction levels are grouped into $2$ \emph{supersets} \cite{marcellin_OnECTCQ} and the occurrences of any level into each superset at each iteration is used for codeword-length estimation. By varying the Lagrange multiplier $\lambda$, the design can fit not only the rate $R$ but any rate up to about $\lceil R\rceil$.
\end{enumerate}

\begin{figure}%
\centering
\subfigure[$\mathbb{Z}/4\mathbb{Z}$-based TCQ]{
\includegraphics[scale=0.6]{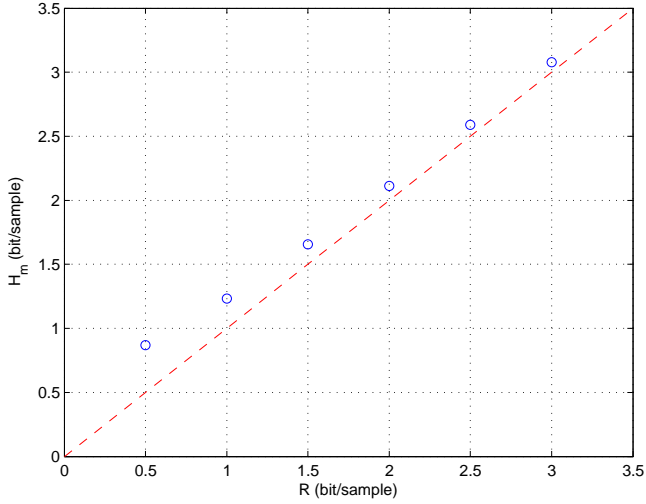}%
\label{f:Hm2R}%
}\\
\subfigure[rate-distortion optimized TCQ]{
\includegraphics[scale=0.6]{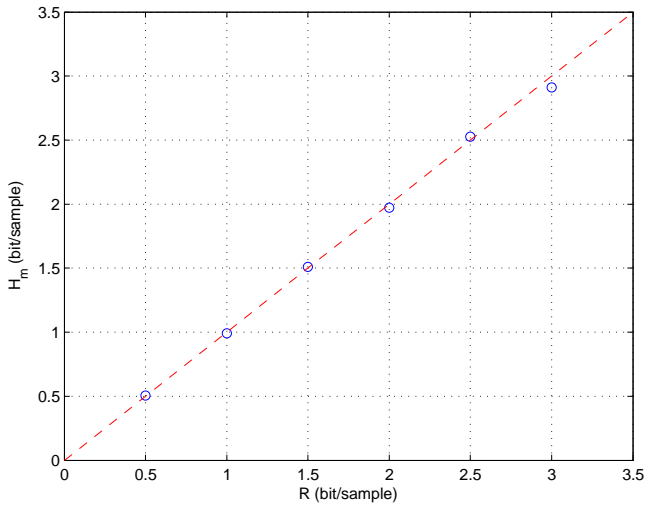}%
\label{f:Hm2R_opt}%
}
\caption{Average entropy vs.~target rate. In case of $\mathbb{Z}/4\mathbb{Z}$-based TCQ (a), the border effects cause the average entropy $H_m$ to be actually greater than the desired target rate $R$; using rate-distortion optimized TCQ (b) it is instead possible to exactly obtain any desired rate.}
\end{figure}

In both cases, a coding system was not actually designed, but the entropy $H$ of the optimal path into the trellis, estimated using the $2$ supersets as contexts, is assumed as achieved rate. On the other hand, it is reasonable that context-based variable-length coding based on the same contexts achieves exactly that rate. Fig.~\ref{f:Hm2R} and Fig.~\ref{f:Hm2R_opt} show the average entropy $H_m$ (in the range BC-SNR~$\in\{9.0,9.5,\dots,19.0\}$ dB, with 8-state syndrome formation) obtained in the two cases in correspondence of the target rates $R\in\{0.5,1.0,\dots,3.0\}$ bit/sample. Using $\mathbb{Z}/4\mathbb{Z}$-based TCQ the entropy is somewhat greater than desired, in particular at $R<2$ bit/sample, due to the impossibility to exactly cover the domain of $s_V(x)$ with polytopes that are translations of the same fundamental Voronoi region. Instead, with rate-distortion optimized TCQ, it is possible to achieve exactly the desired rate, paying the price of a slightly more complex encoder that requires non-uniform quantization and rate-distortion aware metric computation.

\subsection{Finite-Rate Volume Optimization}
In practice, by optimizing the value of $K>1$ it is not only possible to take into account for the non-optimality of the TCQ codes in the case of $s_V(x)$ being correctly received at the decoder, but also possible to take into account for the fact that at low rates $q$ is not negligible w.r.t.~$n$, and hence $P_e\ne 0$. In fact, $K$ is responsible for the normalized volume of the lattice $\Lambda$ and, by increasing its value (at a fixed rate $R$)
\begin{itemize}
  \item the probability of error $P_e\triangleq P[Q_\Lambda(n+q)\ne 0]=P[X - \tilde{s}_V(X)\ne \hat{X} - (\tilde{s}_V(X) + Q)]$ reduces;
  \item the granular error variance $\sigma_q^2$ increases.
\end{itemize}

\begin{figure}%
\centering
\includegraphics[scale=0.6]{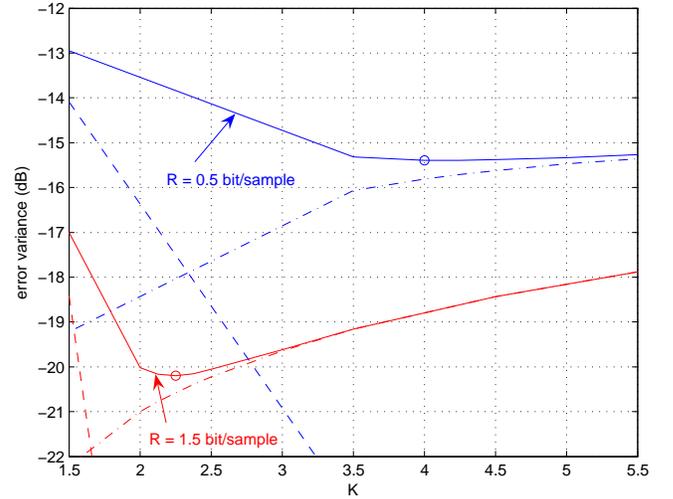}
\caption{Average error variance vs.~volumetric factor, at BC-SNR equal to $14.0$ dB (8-state, $\mathbb{Z}/4\mathbb{Z}$-based TCQ). The solid curves show the effect of varying the volumetric factor $K$ on the final performance at two different target rates $R$. At low $K$, the total error is approximated by the overload error (whose equivalent variance is shown as the dashed curve); at high $K$, the total error is approximated by the granular error (whose equivalent variance is shown as the dot-dashed curve). The circles indicate the optimum values of $K$, which decreases when $R$ increases.}
\label{f:dist2K}
\end{figure}

When an error happens $q_t\simeq q_{ol}$, and hence the total error variance equals
\begin{equation}
\sigma_{q_t}^2\simeq (1-P_e)\sigma_q^2 + P_e\sigma_{q_{ol}}^2\;.\label{e:compose}
\end{equation}
In Fig.~\ref{f:dist2K}, experimental values of both terms of the right hand side of (\ref{e:compose}) are shown in function of $K$ as the dot-dashed and the dashed curve respectively. It is clear that there exist an optimum value of the volumetric factor $K$, possibly different at each rate. The solid curve, i.e.~the experimental value of the total error variance $\sigma_r^2<\sigma_{q_t}^2$ after MSE estimation (with $\sigma_{q_t}^2$ in (\ref{e:MSE_estimate}) hypothesized to equal $\sigma_q^2\simeq K\sigma_n^22^{-2R}$ -- as shown later, the optimum value of $K$ leads indeed to a negligible probability of error $P_e$), shows this effect.

\begin{figure}%
\centering
\subfigure[]{
\includegraphics[scale=0.6]{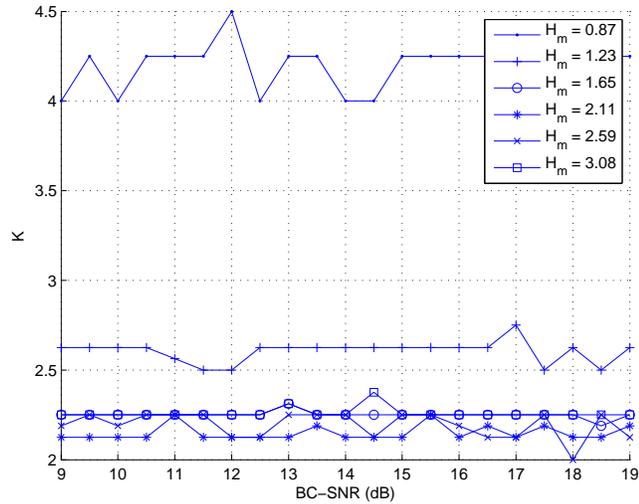}%
\label{f:K2bc_snr}%
}\\
\subfigure[]{
\includegraphics[scale=0.6]{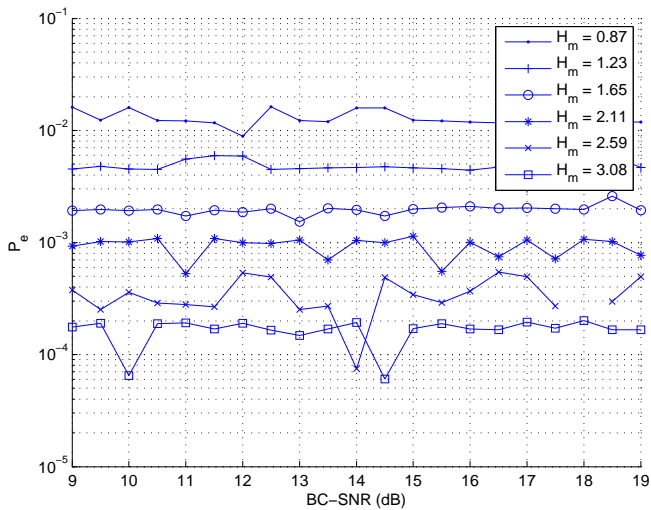}%
\label{f:pe2bc_snr}%
}
\caption{Optimum volumetric factor (a) and corresponding error probability (b) vs.~BC-SNR, at various experimental entropies (8-state, $\mathbb{Z}/4\mathbb{Z}$-based TCQ). }
\end{figure}

For each different code used in syndrome formation, the best value of $K$ was found by a dichotomic-like search at each BC-SNR and at each target rate $R$. The results relative to the 8-state system, reported in Fig.~\ref{f:K2bc_snr}, show that as expected $K$ depends only on the target-rate, and is larger at lower rates, where there would be a significant amount of errors if $K$ was too small. Correspondingly, the optimal value of $P_e$ is constant at the various correlations, as shown in Fig.~\ref{f:pe2bc_snr}.

\begin{figure}%
\centering
\subfigure[]{
\includegraphics[scale=0.6]{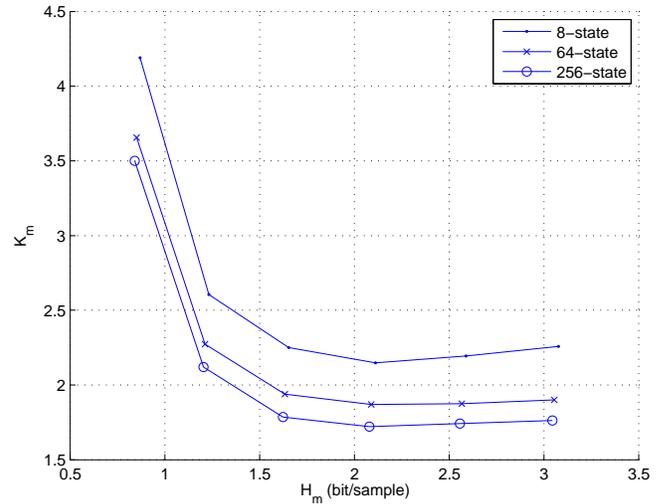}%
\label{f:Km2Hm}%
}\\
\subfigure[]{
\includegraphics[scale=0.6]{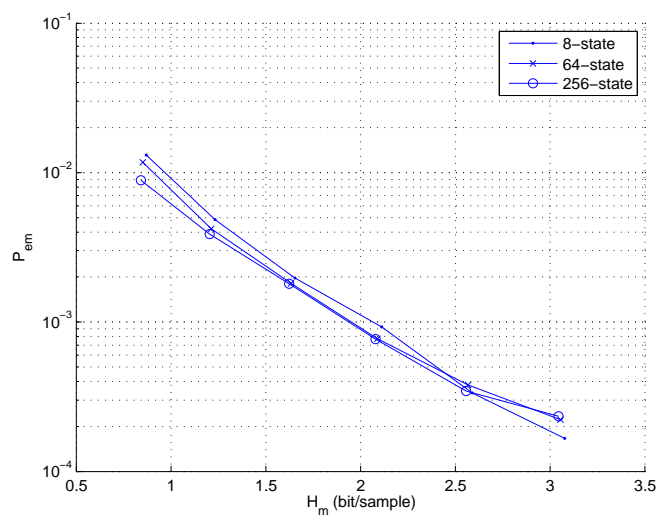}%
\label{f:pe2Hm}%
}
\caption{Average volumetric factor (a) and corresponding average probability of error (b) vs.~experimental entropy, varying the number of states in syndrome formation ($\mathbb{Z}/4\mathbb{Z}$-based TCQ). While $K$ decreases when the number of states increases, i.e.~when the trellis codes are closer to capacity, the probability of error remains almost constant.}
\end{figure}

The average values $K_m$ and $P_{em}$ of $K$ and $P_e$ over the various BC-SNR values are plotted against the average entropy achieved by the encoder in Fig.~\ref{f:Km2Hm} and Fig.~\ref{f:pe2Hm} respectively, using the three different TCQ codes for syndrome formation. Since by increasing the number of states the codes get closer to the capacity of the channel, it is expected, as actually obtained (see Fig.~\ref{f:Km2Hm}), that the optimum values of $K$ reduce; however, the same average probability of error seems to be actually required to achieve the best performance (see Fig.~\ref{f:pe2Hm}). It is interesting to note that the optimal probability of error, which increases at low rates, is very close to the probability of error shown in \cite{pradhan_DISCUS2} in correspondence of the C-SNR that gives the optimal performance in the DISCUS system at both $1$ and $2$ bit/sample.

\subsection{Performance Analysis}
\begin{figure}%
\centering
\includegraphics[scale=0.6]{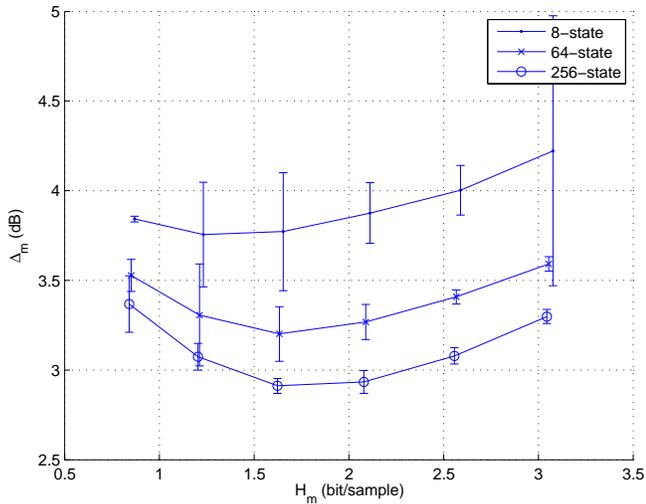}
\caption{Average performance loss (w.r.t.~the Wyner-Ziv bound) vs.~experimental entropy, varying the number of states in syndrome formation ($\mathbb{Z}/4\mathbb{Z}$-based TCQ).}
\label{f:Dm2Hm}
\end{figure}

\begin{figure}%
\centering
\includegraphics[scale=0.6]{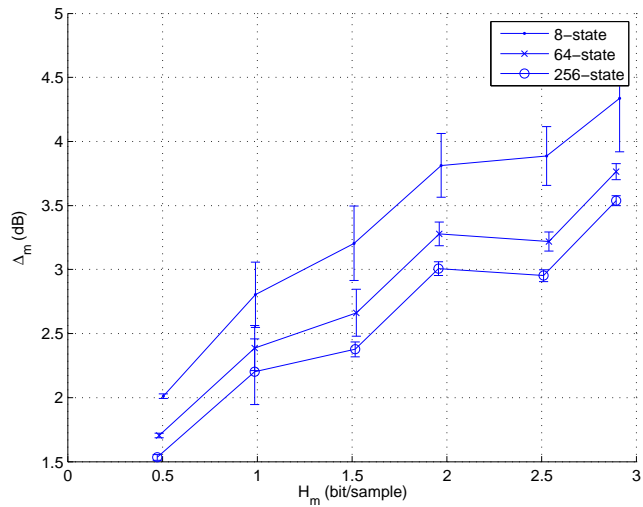}
\caption{Average performance loss (w.r.t.~the Wyner-Ziv bound) vs.~experimental entropy, varying the number of states in syndrome formation (rate-distortion optimized TCQ). Rate-distortion optimized TCQ allows for better performance at low rates (note that in this case the experimental entropy equals the desired target rate).}
\label{f:Dm2Hm_opt}
\end{figure}

The average performance loss w.r.t.~the Wyner-Ziv bound $\Delta_m$ is shown in function of the achieved transmission rate in Fig.~\ref{f:Dm2Hm} and in Fig.~\ref{f:Dm2Hm_opt} for $\mathbb{Z}/4\mathbb{Z}$-based syndrome coding and for rate-distortion optimized syndrome coding respectively. These values were obtained by averaging the experimental value of the performance loss $\Delta\triangleq\sigma_r^2/(\sigma_n^22^{-2H_m})$ over the various values of the BC-SNR in correspondence of the optimum value of $K$. Similarly, the error bars show the average $95$\% confidence intervals\footnote{For each tuple $($BC-SNR$,R,K)$ the $95$\% confidence interval is estimated after up to $5000$ independent simulations.}.

The experiments show that \emph{at any correlation} and in the rate range $R\in[0.5, 3.0]$ bit/sample, the performance of the system is within $3\div 4$ dB of the theoretical bound. In particular, at low rates, it is preferable to code the syndrome using rate-distortion optimized TCQ, which allows for a gain of more than $1$ dB over the other method.

\begin{figure}%
\centering
\subfigure[$R=1$ bit/sample]{
\includegraphics[scale=0.6]{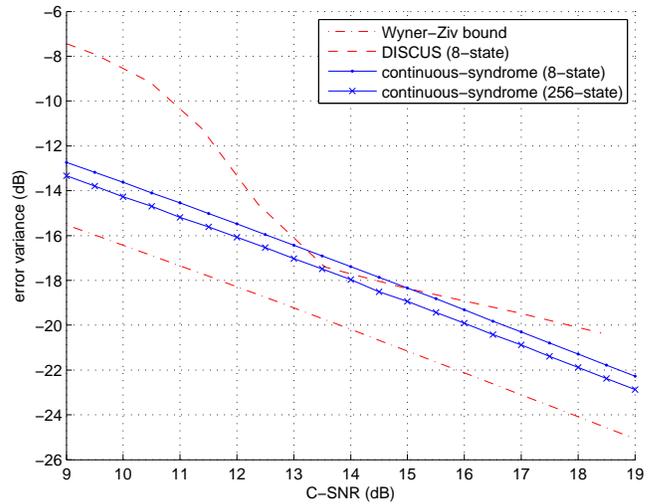}%
\label{f:compare_R1}%
}\\
\subfigure[$R=2$ bit/sample]{
\includegraphics[scale=0.6]{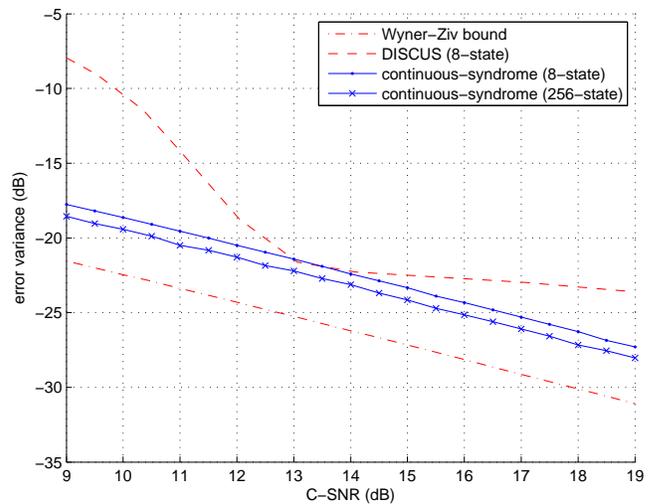}%
\label{f:compare_R2}%
}
\caption{Performance comparison of the proposed algorithm (rate-distortion optimized TCQ, $\sigma_x^2=1$) vs.~the DISCUS algorithm \cite{pradhan_DISCUS2}, at target rate $R=1$ (a) and $R=2$ (b) bit/sample. The proposed algorithm outperforms the DISCUS system, which is optimal for a narrow range of correlations only.}
\end{figure}

To conclude the section, comparisons w.r.t.~to the DISCUS system are shown in Fig.~\ref{f:compare_R1} and  Fig.~\ref{f:compare_R2} for the rates $R=1$ and $R=2$ bit/sample respectively. The immediate result is that by simply choosing the right scaling factor $a$ the proposed system can adapt to any correlation and give the same performance loss, while the DISCUS system must be redesigned to adapt to different correlations. In addition, once the proposed system is designed for a certain additive BCH, no modifications are needed if the statistics of $Y$ (and consequently the one of $X$) changes. Moreover, while only integer rates can be achieved by the DISCUS system, any rate can be achieved by distributed coding based on {\cvs}s. This, again, is simply obtained by choosing the right value of $b$ in case of $\mathbb{Z}/4\mathbb{Z}$-based syndrome coding or, with some increased complexity, by using an ad-hoc rate-distortion optimized TCQ, which finally leads to higher performance at low rates.

While the complexity of the encoding operation is somewhat higher w.r.t.~DISCUS, in which the discrete syndrome is computed on-the-fly after quantization, the decoding complexity is essentially the same, and hence it is perfectly feasible to use distributed coding based on {\cvs}s in actual applications.

\section{Conclusion and Future Work}\label{s:conclusion}
In this paper a coding system with side information at the decoder has been described that operates entirely in the continuous domain. As a consequence of this approach, the system can be designed exactly according to the correlation between the source and the side information and according to the given transmission rate. Extensive experiments showed that the system achieves a good performance in case the correlation can be modeled by a \emph{backward} Gaussian additive channel, even at low rates such as for example $0.5$ bit/sample.

Since in the proposed system the adaptation to the amount of correlation is simply done by means of a scaling factor, it is very interesting to investigate if on-the-fly adaptation to slightly time-varying correlation is possible, which would be of paramount importance in real applications. Moreover, since the actual syndrome coding operation is performed via traditional source coding, it is in principle possible to use scalable source coding tools to obtain a scalable distributed source coding algorithm, that could be applied in frameworks where the available transmission rate is not constant (such as in wireless network-based applications). Both these observations represent a good starting point for future research.

\bibliographystyle{IEEEtran.bst}
\bibliography{../../../IEEEabrv,../../../nonIEEEabrv,../../../refs_TCQ,../../../refs_DSC,../../../refs_books,../../../refs_my}

\end{document}